\documentclass[final,1p,times]{elsarticle} 
\usepackage{graphicx} 
\usepackage{amssymb} 
\usepackage{amsthm} 
\usepackage{lineno} 

\journal{Nuclear Physics A} 
\begin{document} 

\begin{frontmatter} 


\title{Is an Ultra-Cold Strongly Interacting Fermi Gas a Perfect Fluid? }

\author{J. E. Thomas}

\address{Physics Department, Duke University, \\
Durham, NC 27708-0305, USA}

\begin{abstract} 
Fermi gases with magnetically tunable interactions provide a clean
and controllable laboratory system for modeling interparticle
interactions between fermions in nature. The s-wave scattering
length, which is dominant a low temperature, is made to diverge by
tuning near a collisional (Feshbach) resonance. In this regime, two-component
Fermi gases are stable and strongly interacting, enabling tests of
nonperturbative many-body theories in a variety of disciplines,
from high temperature superconductors to neutron matter and
quark-gluon plasmas. We have developed model-independent methods
for measuring the entropy and energy of this model system,
providing a benchmark for calculations of the thermodynamics. Our
experiments on the expansion of rotating strongly interacting
Fermi gases in the normal fluid regime reveal extremely low viscosity
hydrodynamics. Combining the thermodynamic and hydrodynamic
measurements enables an estimate of the ratio of the shear
viscosity to the entropy density. A strongly interacting Fermi gas
in the normal fluid regime is found to be a nearly perfect fluid,
where the ratio of the viscosity to the entropy density is close
to a universal minimum that has been conjectured by string theory
methods.
\end{abstract} 

\end{frontmatter} 



\section{Introduction}

Tabletop experiments with degenerate atomic Fermi gases near a
Feshbach resonance~\cite{OHaraScience,AmScientist} provide models
for strongly interacting Fermi systems in nature. Feshbach
resonances~\cite{HoubiersSF,Houbiers12} arise when different
hyperfine channels have different magnetic moments, as occurs in
$^6$Li and $^{40}$K atomic Fermi gases. A bias magnetic field is
applied to tune the total collision energy of the incoming
continuum state into resonance with that of a bound state in a
closed channel. At resonance, the zero energy s-wave scattering
length $a_S$ diverges, and the collision cross section is only limited
by unitarity, i.e., $\sigma\propto \lambda_B^2$, where $\lambda_B$
is the de Broglie wavelength. Even though it is a dilute system, a
unitary atomic gas is the most strongly interacting
non-relativistic system known~\cite{SchaferViscosity}.

Strongly interacting Fermi gases exhibit strong pairing interactions,
 of interest in the field of high temperature superconductivity~\cite{Levin}, neutron
stars, and nuclear matter~\cite{Bertsch,Baker,Heiselberg,Carlson}. The common feature which all of these
systems share is a strong interaction between pairs of spin-up and
spin-down fermions. The strongly collisional normal
fluid exhibits extremely low viscosity hydrodynamics and elliptic flow~\cite{OHaraScience}, analogous to the hydrodynamics of a quark-gluon
plasma~\cite{Heinz,Shuryak}.  In contrast to other Fermi systems, atomic
gases enable magnetically tunable
interactions~\cite{Grimmbeta,GrimmLeeHuangYang,JosephSound},
variable energy~\cite{JointScience,KinastDampTemp,LuoEntropy}, and
variable spin populations~\cite{KetterleImbalanced,HuletPhaseSeparation}.
Fermi gases near Feshbach resonances are now being widely studied~\cite{RMP2008}.

It has been suggested that a strongly interacting Fermi gas may
serve as a model for the most ``perfect" fluid, which has a
minimum viscosity. A simple argument based on quantum mechanics
places a lower bound on the viscosity~\cite{GyulassyViscosity}.
The shear viscosity is of order $\eta\propto n\,p\lambda_{mfp}$,
where $n$ is the density, $p$ is the average momentum and
$\lambda_{mfp}$ is the mean free path~\cite{Kittel}. The
Heisenberg uncertainty principle requires
$p\lambda_{mfp}\geq\hbar$, so that the shear viscosity satisfies
$\eta\geq n\,\hbar$. In high energy physics, where particle number
is not conserved, the ratio of the viscosity to the entropy
density $s\simeq n\,k_B$ is often considered. Hence, one expects
that $\eta/s \geq \hbar/k_B$.

Recently, using string theory methods, it has been conjectured
that there is a universal strong coupling
lower bound~\cite{KovtunViscosity},
\begin{equation}
\frac{\eta}{s}\geq\frac{1}{4\pi}\frac{\hbar}{k_B}.
\label{eq:lowerbound}
\end{equation}
Although one can imagine high entropy systems
which might violate the lower bound~\cite{CohenViscosity},
currently no fluid that even achieves the lower bound
is known. If the viscosity conjecture is correct, it represents an
extremely important advance in the understanding of many-body
physics~\cite{CohenViscosity}. Hence, it is of great interest to
explore minimum viscosity quantum hydrodynamics in strongly
interacting Fermi gases as a model
system~\cite{SchaferViscosity,TurlapovPerfect}.

An important feature of strongly interacting Fermi gases is the
property of
universality~\cite{Heiselberg,HoUniversalThermo,ThomasUniversal,DrummondUniversal}.
The system exhibits scale invariance, in the sense that, at zero temperature, the
interparticle spacing $L$ sets the only microscopic length scale at
resonance, leading to universal behavior.  In a uniform strongly
interacting gas, the ground-state energy is a universal fraction,
denoted $1+\beta$, of the energy of a noninteracting gas at the
same density~\cite{OHaraScience,MechStab}. This universal energy
relationship was originally explored theoretically in the context
of nuclear matter~\cite{Bertsch,Baker,Heiselberg,Carlson} and has
now been measured using ultracold Fermi
atoms~\cite{OHaraScience,Grimmbeta,JosephSound,MechStab,SalomonExpInt,JinMomentum}.
Our best current measurements of $\beta$ are given in Ref.~\cite{Thermo09}.

Universality automatically leads to the natural
``quantum viscosity scale" $\hbar\,n$~\cite{ShuryakQuantumViscosity}. The shear viscosity has natural units of momentum/area. In a strongly interacting Fermi gas,
the natural momentum is $\hbar/L$, while, the natural area is the unitary collision cross section, of order $L^2$ for temperatures at or below the Fermi temperature. The shear viscosity is then of order $\hbar/L^3$ or $\hbar\,n$.  It is therefore natural to write the viscosity in the form,
\begin{equation}
\eta = \alpha\,\hbar\,n,
\label{eq:viscosity}
\end{equation}
where $\alpha$ is a dimensionless parameter, which is generally a function of the local reduced temperature, $T/T_F(n)$, where $T_F(n)$ is the local Fermi temperature, $\propto n^{2/3}$. We note that for water $\alpha \simeq 300$, while for air, $\alpha \simeq 6000$. For liquid He near the $\lambda$-point, $\alpha\simeq 1$, in the quantum regime. As we will see, our estimates for a strongly interacting Fermi gas are significantly lower.

For a trapped gas, we are able to estimate the ratio $\eta /s$ as~\cite{TurlapovPerfect}
\begin{equation}
\frac{\eta}{s}=\frac{\hbar}{k_B}\frac{\langle \alpha\rangle}{S/k_B},
\label{eq:etasratio}
\end{equation}
where $S/k_B$ is the average entropy per particle for the trapped gas and $\langle\alpha\rangle$ is the trap averaged shear viscosity in units of $\hbar n$.
In the following, we describe our recent studies of the thermodynamics, i.e., the measurement of the energy and entropy. Then we will describe our studies of the hydrodynamics and the estimate of the shear viscosity. Using these results, we compare to the lower bound of Eq.~\ref{eq:lowerbound}.

\section{Measuring the Entropy and Energy}

\begin{figure}[htb]
\begin{center}
\includegraphics[width=2.0in]{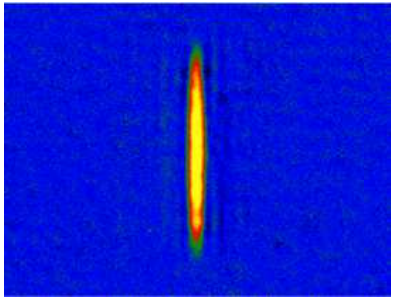}
\end{center}
\caption{Absorption image of an optically trapped, strongly interacting Fermi gas. The length of the cloud is $\simeq 200\,\mu$m. }
 \label{fig:cloud}
 \end{figure}

Recently, we have developed model independent methods for measuring the entropy and energy of a strongly interacting Fermi gas~\cite{LuoEntropy,Thermo09}.
Energy measurement is based on the virial theorem~\cite{ThomasUniversal}. Since the interparticle spacing and the thermal de Broglie wavelength are the only length scales when the cloud is tuned to a broad Feshbach resonance, the local pressure is a function only of the local density and temperature. In this case, one easily verifies that the virial theorem holds using elementary thermodynamic arguments, which is confirmed by experiment~\cite{ThomasUniversal}. The atoms are confined in an optical trap, which at low temperatures, provides a nearly harmonic trapping potential, yielding the energy

\begin{equation}
E = 2\langle U\rangle = 3 m\omega_z^2\langle z^2\rangle_{S} .
\label{eq:energy}
\end{equation}
Here, we have assumed a scalar pressure, so that the harmonic trapping potential energy is identical in all three directions, $x,y,z$. We make measurements of the mean square size in the long $z$-direction of the cigar-shaped cloud, Fig.~\ref{fig:cloud}. The spring constant, $m\omega_z^2$ is typically determined within 0.5\%, and the mean square size of the strongly interacting gas $\langle z^2\rangle_{S}$ is determined within 2\%.
Despite the fact that the gas generally contains condensed superfluid pairs, non-condensed pairs, and unpaired atoms, all strongly interacting, this remarkable result shows that a simple measurement of the mean square size of the trapped cloud determines the total energy.

The entropy is measured by means of an adiabatic sweep of the bias magnetic field, to tune the scattering length $a_S$ from the strongly interacting regime,  $na_S^3>>1$, to the weakly interacting regime, where $na_S^3<<1$. In the weakly interacting regime, the entropy $S_W$ of the cloud is nearly that of an ideal Fermi gas in a harmonic trap, where
\begin{equation}
S_W\simeq S_{ideal}=\int d\epsilon\, {\cal D}(\epsilon)\,s(\epsilon,T).
\label{eq:entropy}
\end{equation}
Here $s(\epsilon,T)$ is the Boltzmann entropy for an orbital of single particle energy $\epsilon$ at temperature $T$, and ${\cal D}(\epsilon)$ is the density of states for the trap. The chemical potential is determined by normalizing the total occupation number to the total number of atoms. Using this, the entropy, energy and mean square size of the cloud are then determined for a  given temperature $T$.  Eliminating the temperature, the entropy is given in terms of the mean square cloud size of the weakly interacting gas $\langle z^2\rangle_W$, which is readily measured,
\begin{equation}
S_W\simeq S_{ideal}\left(\langle z^2\rangle_W -\langle z^2\rangle_{W0}\right).
\label{eq:entropy2}
\end{equation}
Here, $\langle z^2\rangle_{W0}$ is the mean square size of the cloud in the ground state, which is estimated by fitting  a Thomas-Fermi density profile to the density profiles at the lowest temperature, yielding the Fermi radius, $\sigma_z$, from which $\langle z^2\rangle_{W0}=\sigma_z^2/8$. The measured value is in very good agreement with calculations  based on the local chemical potential at zero temperature $\mu_0(n)$~\cite{Thermo09}. We note that writing the entropy as a function of the mean square size relative to the ground state assures $S_W=0$ for the measured ground state cloud size. This method also improves the ideal gas approximation by suppressing small mean field corrections to the ideal gas cloud sizes at finite interaction strength. We find that the corrections to the ideal gas entropy for the finite interaction strength are within a few percent except at our lowest temperatures, where the correction is $\simeq 10$\%~\cite{Thermo09}.

 To verify that the sweep of the bias magnetic field is adiabatic, we note that  after a round trip sweep lasting 2 s between the strongly and weakly interacting regimes, the energy is found to be within 2\% of that obtained by simply holding the strongly interacting cloud for 2 s. Hence,
 \begin{equation}
 S_S=S_W .
 \end{equation}

\begin{figure}
\centerline{\includegraphics[width= 3.5in,clip]{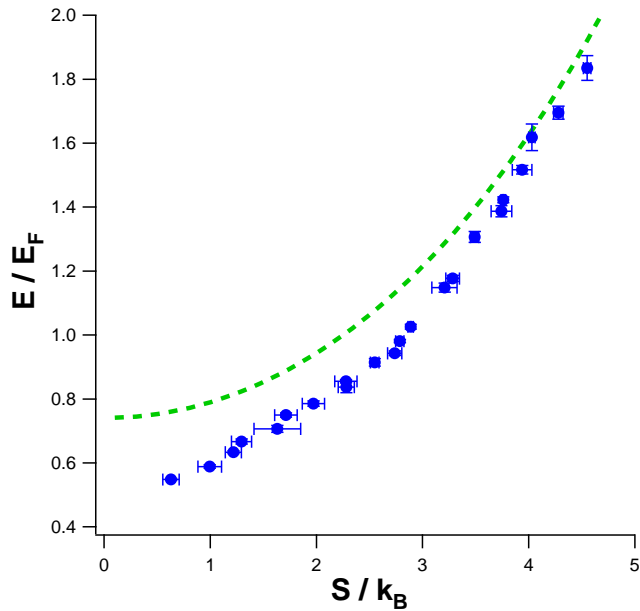}}
\caption {Measured total energy per particle in units of $E_F$ of
a strongly interacting Fermi gas at 840 G versus its entropy per
particle in units of $k_B$. For comparison, the dot-dashed green
curve shows $E(S)$ for an ideal Fermi gas. $E_F$ is the Fermi
energy of an ideal Fermi gas at the trap center.
\label{fig:energyentropy}}
\end{figure}

 To perform the measurements, an atom cloud is first cooled by lowering the trap depth to achieve forced evaporation to a temperature near the ground state. Then energy is added by releasing the cloud for a precisely controlled time and then recapturing the cloud. This method reproducibly adds energy to the cloud, which is allowed to equilibrate. The energy of the strongly interacting cloud is then measured from the cloud size. A second cloud is then created using the same parameters as the first. The bias magnetic field is swept over 1 s to the weakly interacting regime, where the mean square size is measured. Together, these measurements yield the energy and entropy of the strongly interacting gas, Fig.~\ref{fig:energyentropy}.

The scaling of the energy with entropy is quite different for low energies  $E/E_F<0.8$ than for higher energies $E/E_F>0.8$. We attribute this change in the thermodynamics to a superfluid transition. Fitting a smooth curve to the data, and using $T=\partial E/\partial S$, we find that superfluid-normal fluid transition occurs at a temperature $T\simeq 0.2\,T_F$~\cite{Thermo09}, which is quite high. In a charged condensed matter system, where $T_F$ corresponds to an eV or $10^4$ K, this would correspond to a superconductor transition at 2000 K!

\section{Estimating the Shear Viscosity}

For a unitary Fermi gas, the bulk viscosity is believed to vanish~\cite{SonBulkViscosity,EscobedoBulkViscosity}, so that the shear viscosity determines the hydrodynamic damping rate. Our first estimates of the shear viscosity were made by measuring  the damping rate of the radial breathing mode~\cite{TurlapovPerfect}. Sch\"{a}fer~\cite{SchaferViscosity} used this damping data to estimate the shear viscosity and combined the results with our entropy data to make a comparison with the lower bound of Eq.~\ref{eq:lowerbound}. We noted previously that edge effects in the trap, such as interactions between the hydrodynamic and ballistic components of the cloud, might have increased the observed damping rate in the trapped gas. The damping rates do not appear to scale properly with atom number at fixed $E/E_F$ if viscosity is assumed to be the primary cause of damping~\cite{TurlapovPerfect}. However, the estimated viscosity is in the quantum regime.

\begin{figure*}[tb]
\includegraphics[height=2.25 in]{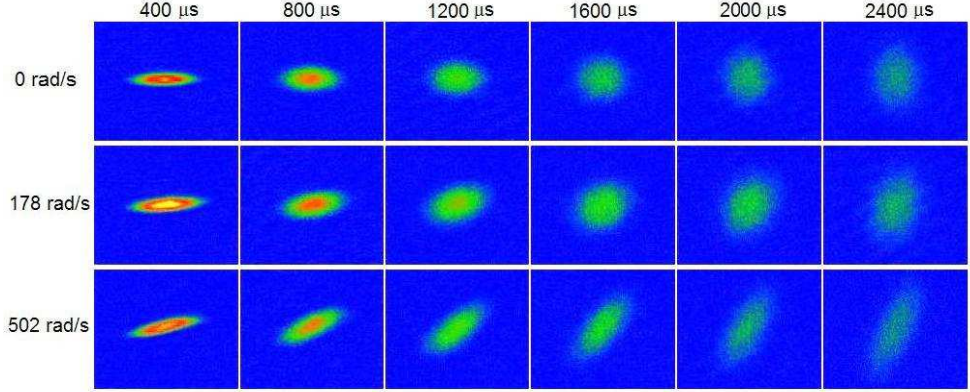}
\caption{Hydrodynamic expansion of a rotating, strongly interacting Fermi gas, for different initial angular velocities.} \label{fig:rotating}
\end{figure*}

Recently, we have estimated the shear viscosity in a different way, by measuring the expansion dynamics of a rotating Fermi
gas, which is released from the optical trap~\cite{ClancyRot07}. In this case, the gas expands freely, eliminating the direct effects of the edges of a trapped cloud. The gas is cooled by evaporation to near the
ground state and a controlled amount of energy is added. Then the
trap is rotated abruptly to excite a scissors mode. The
gas is released and imaged after a selected expansion time.
Fig.~\ref{fig:rotating} shows typical data, for different initial angular
velocities. The angle of the long principal axis of the cloud is measured as a function of time after release. The data reveal that as the gas expands, the angular velocity increases, which is a consequence of irrotational hydrodynamics: The moment of inertia
decreases as the aspect ratio approaches unity.

Remarkably, the cloud for the normal fluid at $E/E_F=2.1$  behaves
almost identically to the superfluid cloud for $E/E_F=0.56$.
Indeed, the effective moment of inertia is quenched well below the rigid body
value in both cases, and is in very good agreement with
expectations for irrotational flow\cite{ClancyRot07}.

Irrotational flow is expected for the superfluid, since the
velocity field is the gradient of the phase of a macroscopic
wavefunction. However, irrotational flow in the normal fluid requires very
low shear viscosity. As energy is added to the gas, we find that the expansion dynamics slows down compared to ideal, isoentropic irrotational flow.

\begin{figure}[htb]
\includegraphics[height=2.5in]{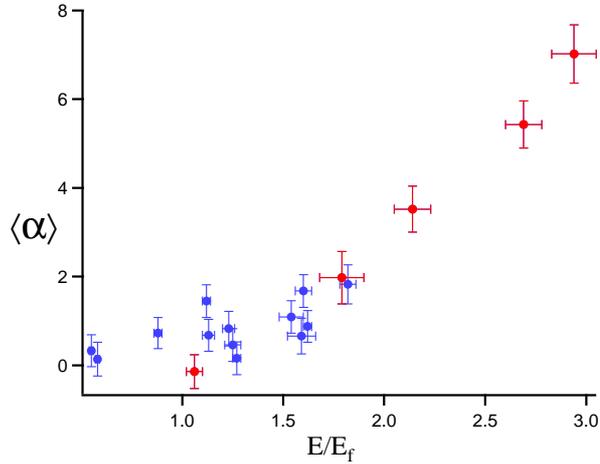}
\caption{Estimated shear viscosity in units of $\hbar\,n$ versus energy in
units of $E_F$. Red (blue) circles denote data at 20\% (5\%) of maximum trap depth.} \label{fig:alpha}
\end{figure}

To estimate the shear viscosity, we use a simple model. We assume that the slowing down of the dynamics, compared to ideal irrotational flow, arises from shear viscosity. In the absence of viscosity, the
expansion is isoentropic. In this case, for release from a harmonic trap, an exact solution to the hydrodynamic equations is obtained
for a velocity field that is linear in the spatial coordinates~\cite{ClancyRot07}. We then add to the hydrodynamic equations a term that is the divergence of the pressure tensor arising from shear viscosity~\cite{ClancyThesis08}. The time evolution depends on the trap average $\langle\alpha\rangle$, Eq.~\ref{eq:viscosity}, which increases as the energy is increased. At energies just above the superfluid transition, the viscosity is sufficiently small that the uncertainty is determined by the accuracy of the trap parameters (oscillation frequencies in the three directions are determined within 0.5\%). Note that for the lowest viscosities, the fit can return a negative value of $\langle\alpha\rangle$, which is an artifact arising from measured trap parameters that predict a slightly slower evolution for perfect irrotational flow than that measured.  Fig.~\ref{fig:alpha} shows how the estimated shear viscosity depends on the energy of the cloud. The shear viscosity is given in units of the quantum
viscosity, i.e., in units of $\hbar \,n$, where $n$ is the density. The red (blue) data are taken at trap depths that are 20\% (5\%) of the maximum attainable. The agreement in the data for both trap depths shows that anharmonicity in the Gaussian trapping potential is not significant.

\begin{figure}[htb]
\includegraphics[height=2.5in]{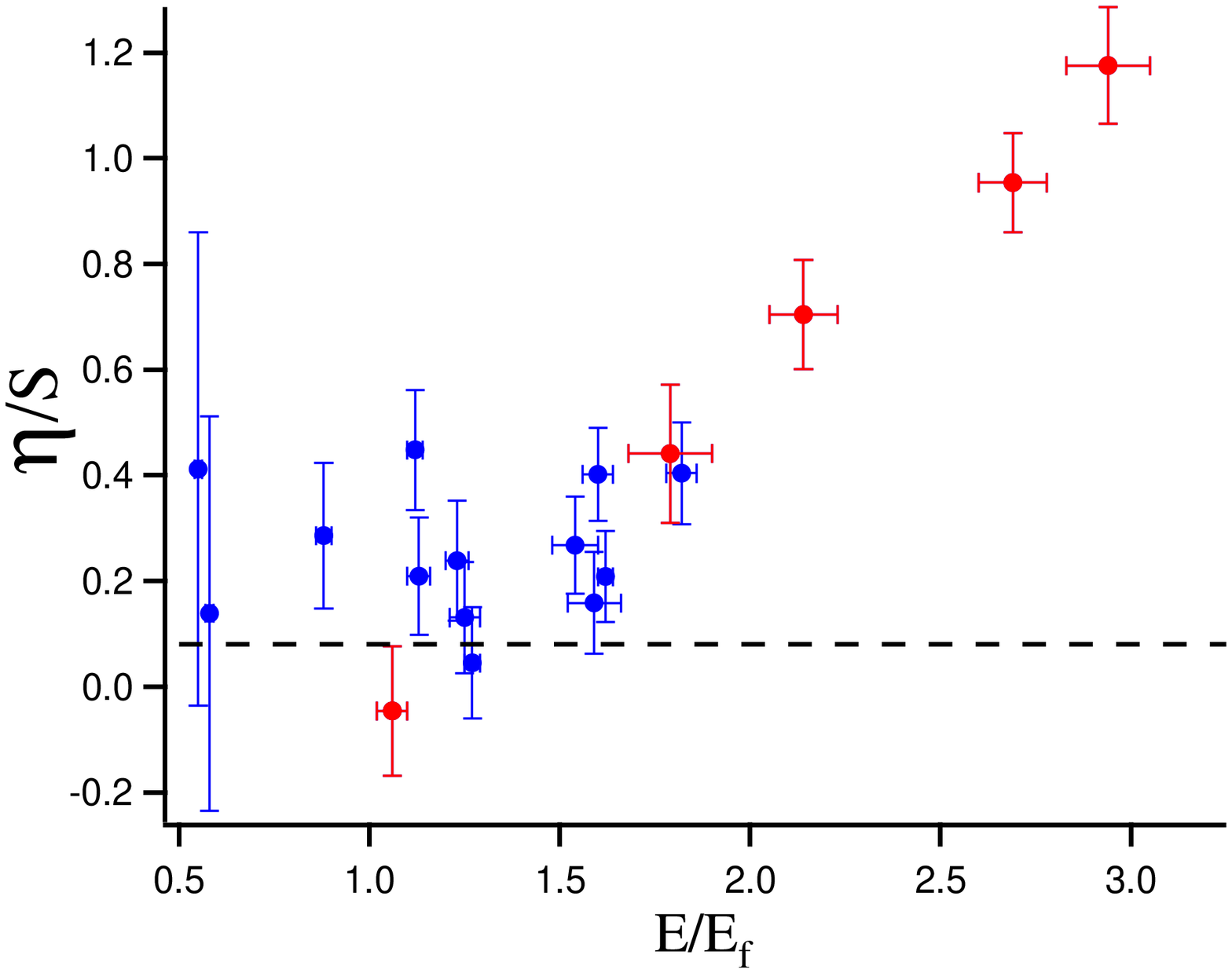}
\caption{Estimated ratio of the shear viscosity to the entropy
density. Dotted line shows the string theory conjecture~\cite{KovtunViscosity} for the
minimum ratio, Eq.~\ref{eq:lowerbound}.} \label{fig:etas}
\end{figure}

By combining the entropy measurements with the viscosity estimates, we are able
to estimate the ratio of the shear viscosity to entropy density.
Fig.~\ref{fig:etas} shows how the results  compare to the string theory
conjecture for the minimum ratio, Eq.~\ref{eq:lowerbound}. Our estimates of the viscosity suggest that a strongly interacting Fermi gas
in the normal fluid regime (above $0.8\,E_F$) is a nearly perfect
fluid.

\section*{Acknowledgments}
This research was supported by the Physics Divisions of the National Science Foundation and the Army
Research Office, and the Chemical Sciences, Geosciences and Biosciences Division of  the
Office of Basic Energy Sciences, Office of Science, U.S. Department
of Energy. I am indebted to  Bason Clancy, Le Luo, James Joseph, Chenglin Cao, and Jessie Petricka, who performed the experiments
and analysis reported in this paper.



\end{document}